\documentclass[aps, pra, reprint, lengthcheck, footinbib, showpacs,  citeautoscript]{revtex4-2}

\usepackage[T1]{fontenc}

\usepackage{multirow}

\usepackage{amsmath}

\usepackage{amsmath}

\usepackage{braket}
\usepackage{graphicx}

\usepackage[colorlinks, urlcolor=blue, citecolor=blue, linkcolor=blue, pdfstartview=FitH]{hyperref}
\usepackage[all]{hypcap}

\usepackage{dcolumn}
\usepackage{bm}
\usepackage{epstopdf}

\usepackage{xcolor}
\usepackage[normalem]{ulem}
\definecolor{color}{rgb}{0.11,0.45,0.02}

\begin{document}

\title{Hyperfine interaction of electrons confined in CsPbI$_3$ nanocrystals with nuclear spin fluctuations 
}

\author{Sergey~R.~Meliakov$^{1}$, Evgeny~A.~Zhukov$^{2,1}$, Vasilii~V.~Belykh$^{2}$, Kirill V. Kavokin$^{3}$, Mikhail O. Nestoklon$^{2}$, Elena~V.~Kolobkova$^{4,5}$, Maria~S.~Kuznetsova$^3$, Manfred Bayer$^{2,6}$, Dmitri~R.~Yakovlev$^{2,1}$}

\affiliation{$^{1}$P.N. Lebedev Physical Institute of the Russian Academy of Sciences, 119991 Moscow, Russia}
\affiliation{$^{2}$Experimentelle Physik 2, Technische Universit\"at Dortmund, 44227 Dortmund, Germany}
\affiliation{$^{3}$Spin Optics Laboratory, St. Petersburg State University, 198504 St. Petersburg, Russia}
\affiliation{$^{4}$ITMO University, 199034 St. Petersburg, Russia}
\affiliation{$^{5}$St. Petersburg State Institute of Technology, 190013 St. Petersburg, Russia}
\affiliation{$^6$Research Center FEMS, Technische Universit\"at Dortmund, 44227 Dortmund, Germany}

\date{\today}

\begin{abstract}

The coherent spin dynamics of electrons are investigated for CsPbI$_3$ perovskite nanocrystals in a glass matrix using time-resolved Faraday ellipticity. In nanocrystals with a diameter of about 11~nm, the Larmor precession frequency has a linear dependence on magnetic field corresponding to the electron Land\'e $g$-factor of 2.07. We find a finite Larmor precession frequency at zero magnetic field, corresponding to the electron spin splitting of $0.8$~$\mu$eV. This splitting is explained by the hyperfine interaction with nuclear spin fluctuations. Our model analysis shows that the hyperfine interaction for the conduction band electrons is contributed both by the $p$-orbitals of the lead atoms and by the $s$-orbitals of the iodine atoms, with the leading contribution to the hyperfine field fluctuations coming from iodine. This fact agrees well with the 9\% iodine contribution to the Bloch amplitude of the conduction band, obtained by DFT calculations. From these findings, the atomic hyperfine constant for the $5s$-orbital of iodine is evaluated as 190 $\mu$eV.      

\end{abstract}

\maketitle



\section{Introduction}

Semiconductors based on lead halide perovskite compounds are actively studied nowadays by physicists, chemists and engineers. The interest is inspired by their photovoltaic and optoelectronic properties~\cite{Jena2019, Vinattieri2021_book, Vardeny2022_book}. They  are also highly interesting for basic research, as their band structure differs considerably from the one of common III-V and II-VI semiconductors, which allows one to examine new settings for the emergence of novel phenomena.  The possibility to have these materials not only as bulk single crystals, but also as two-dimensional layers and nanocrystals (NCs) provides quantum confinement as tool for tuning the material properties and searching for nanostructure-specific effects.   



Furthermore, perovskite semiconductors show attractive spin properties, which can be accessed by the optical and magneto-optical experimental techniques that have been well-established in the spin physics of semiconductors~\cite{OOBook1984,SpinBook2017}. Among them, there is time-resolved Faraday/Kerr rotation (TRFR or TRKR), which allows one to address the coherent spin dynamics of electrons and holes, and to measure their spin coherence, spin dephasing, and longitudinal spin relaxation times, as well as their Land\'e $g$-factors~\cite{belykh2019,kirstein2022nc,kirstein2022am}. Elaborated experimental protocols using two-color and/or two-pump techniques allow one to implement all-optical spin control~\cite{Lin2023,Zhukov_2025two_colors}, resonant spin amplification and accumulation~\cite{Kirstein_2025RSAmapi}, and mode-locking of hole spin coherence~\cite{kirstein2023_SML}. TRFR was successfully used to study the spin properties of all-inorganic CsPbBr$_3$ and CsPbI$_3$ NCs, synthesized by colloidal chemistry in solution~\cite{Crane2020,Grigoryev2021,Lin2023,Gao2024,Gao2024NL,Zhu2024,Cheng2024,Cai2023}, as well as fabricated in a glass matrix~\cite{kirstein2023_SML,nestoklon2023_nl,Meliakov2024_paper1,Meliakov2025_paper2,Meliakov2024_paper3}.  




The spin-dependent phenomena in semiconductors include the interaction of the carrier spins with the nuclear spins. Cryogenic temperatures and strong carrier localization in, e.g., quantum dots (QDs) greatly enhance the manifestations of this interaction~\cite{Stepanenko2006,Chekhovich2013,Bechtold2015,Abobeih2018}. The nuclear spins can be dynamically polarized via their interaction with optically-oriented carriers. In turn, the Overhauser field of the polarized nuclei acts back on the carrier spins, causing their spin splitting and facilitating their spin dephasing. Dynamic nuclear polarization was reported for several perovskite crystals, namely: FA$_{0.9}$Cs$_{0.1}$PbI$_{2.8}$Br$_{0.2}$~\cite{kirstein2022am}, MAPbI$_3$~\cite{kirstein2022mapi}, FAPbBr$_3$~\cite{Kirstein2023DNSS}, and CsPbBr$_3$~\cite{belykh2019}. 

Even in case of unpolarized nuclei, their spin fluctuations may influence the spin dynamics of localized carriers. This problem was theoretically considered in 2002 by Merkulov, Efros and Rosen~\cite{Merkulov2002} for III-V and II-VI semiconductor quantum dots, where the electrons in the conduction band with $s$-type wave function interact with the nuclei much stronger than the holes in the valence band with $p$-type wave function. In lead halide perovskites, the valence band has significant contribution from the $s$-orbitals of the Pb-ions, while the conduction band is formed mainly by $p$-orbitals of the Pb-ions with a small admixture of $s$-orbitals of the halogens. As a result, the hole-nuclei hyperfine interaction is about 5 times stronger than the electron-nuclei one~\cite{kirstein2022am}.

The role of the nuclear spin fluctuations in the spin dynamics of localized electrons and holes was studied for FA$_{0.9}$Cs$_{0.1}$PbI$_{2.8}$Br$_{0.2}$ perovskite crystals by means of the Hanle and polarization recovery effects~\cite{kudlacik2024_oo_carriers}. In FAPbBr$_3$ crystals, the suppression of the nuclear spin fluctuations via creation of a squeezed dark nuclear spin state was demonstrated~\cite{Kirstein2023DNSS}.  Very recently, we reported on the hyperfine interaction between the holes and the nuclear spin fluctuations in CsPbBr$_3$ and CsPb(Cl,Br)$_3$ NCs in glass~\cite{Meliakov2024_paper3}.


In this paper, time-resolved Faraday ellipticity is used to study the coherent spin dynamics of electrons in CsPbI$_3$ NCs embedded in a glass matrix. At zero magnetic field, electron spin precession in the random hyperfine fields of the nuclear spin fluctuations is detected. A comparison of the observed oscillations with the Merkulov-Efros-Rosen theory yields the dispersion of the hyperfine field, which is then used to estimate the contribution of the iodine $s$-orbitals to the conduction-band Bloch amplitude in CsPbI$_3$. The experimental value shows good agreement with the results of our DFT calculations. Using the DFT results, we were able to determine the atomic hyperfine constant for the $5s$-shell of iodine, which so far has not been measured for iodine atoms in solids.  

\section{Experimental details}

The studied CsPbI$_3$ nanocrystals embedded in fluorophosphate Ba(PO$_3$)$_2$-AlF$_3$ glass were synthesized by rapid cooling of a glass melt enriched with the components needed for perovskite crystallization. Details of the method are given in Refs.~\onlinecite{Kolobkova2021,kirstein2023_SML}. Samples of fluorophosphate (FP) glass with the composition 35P$_2$O$_5$--35BaO--5AlF$_3$--10Ga$_2$O$_3$--10PbF$_2$--5Cs$_2$O (mol. \%) doped with BaI$_2$ were synthesized using the melt-quench technique. The glass synthesis was performed in a closed glassy carbon crucible at the temperature of $T=1050^\circ$C. About 50~g of the batch was melted in the crucible for 30~min., then the glass melt was cast on a glassy carbon plate and pressed to form a plate with a thickness of about 2~mm. Samples with a diameter of 5~cm were annealed at the temperature of $50^\circ$C below $T_g=400^\circ$C to remove residual stresses. The CsPbI$_3$ perovskite NCs were formed from the glass melt during the quenching. The glasses obtained in this way are doped with CsPbI$_3$ NCs. The dimensions of the NCs in the initial glass were regulated by the concentration of iodide and the rate of cooling of the melt without heat treatment above $T_g$. The technology code of the sample is EK8. The studied sample has a dispersion of NCs size from $9$~nm to $12$~nm.

To study the coherent spin dynamics of electrons we use a pump-probe time-resolved technique with detection of the Faraday ellipticity (TRFE)~\cite{Yakovlev_Ch6,Glazov2010}.  Spin oriented electrons are generated by the circularly polarized pump pulses. The used laser system (Light Conversion) has two optical parametric oscillators (OPA) generating pulses of either 1.5 ps duration (with spectral width of about 1~meV) or 150~fs duration (spectral width of about 20~meV) at the repetition rate of 25~kHz (repetition period 40~${\mu}$s). The laser photon energy ($E_{\rm L}$) can be tuned in the spectral range of $1.65-1.85$~eV in order to resonantly excite NCs of various sizes. The laser beam is split into the pump and probe beams, whose photon energies coincide. The time delay between the pump and probe pulses is controlled by a mechanical delay line. The pump beam is modulated with an electro-optical modulator between ${\sigma}^+$ and ${\sigma}^-$ circular polarization at the frequency of 26~kHz. The probe beam is linearly polarized. The Faraday ellipticity of the probe beam, which is proportional to the electron spin polarization, is measured as function of the delay between the pump and probe pulses using a balanced photodetector connected to a lock-in amplifier that is synchronized with the modulator. Both pump and probe beams have the power of 0.5~mW and spot sizes of about 100$~\mu$m. The sample is placed on the cold finger of a helium-flow optical cryostat at the temperature of $T=6$~K. A magnetic field up to 430~mT is generated by an electromagnet and applied perpendicular to the laser $k$-vector (Voigt geometry, $\textbf{B} \perp \textbf{k}$).

\section{Experimental results}

\begin{figure}[bt!]
\includegraphics[width=1\columnwidth]{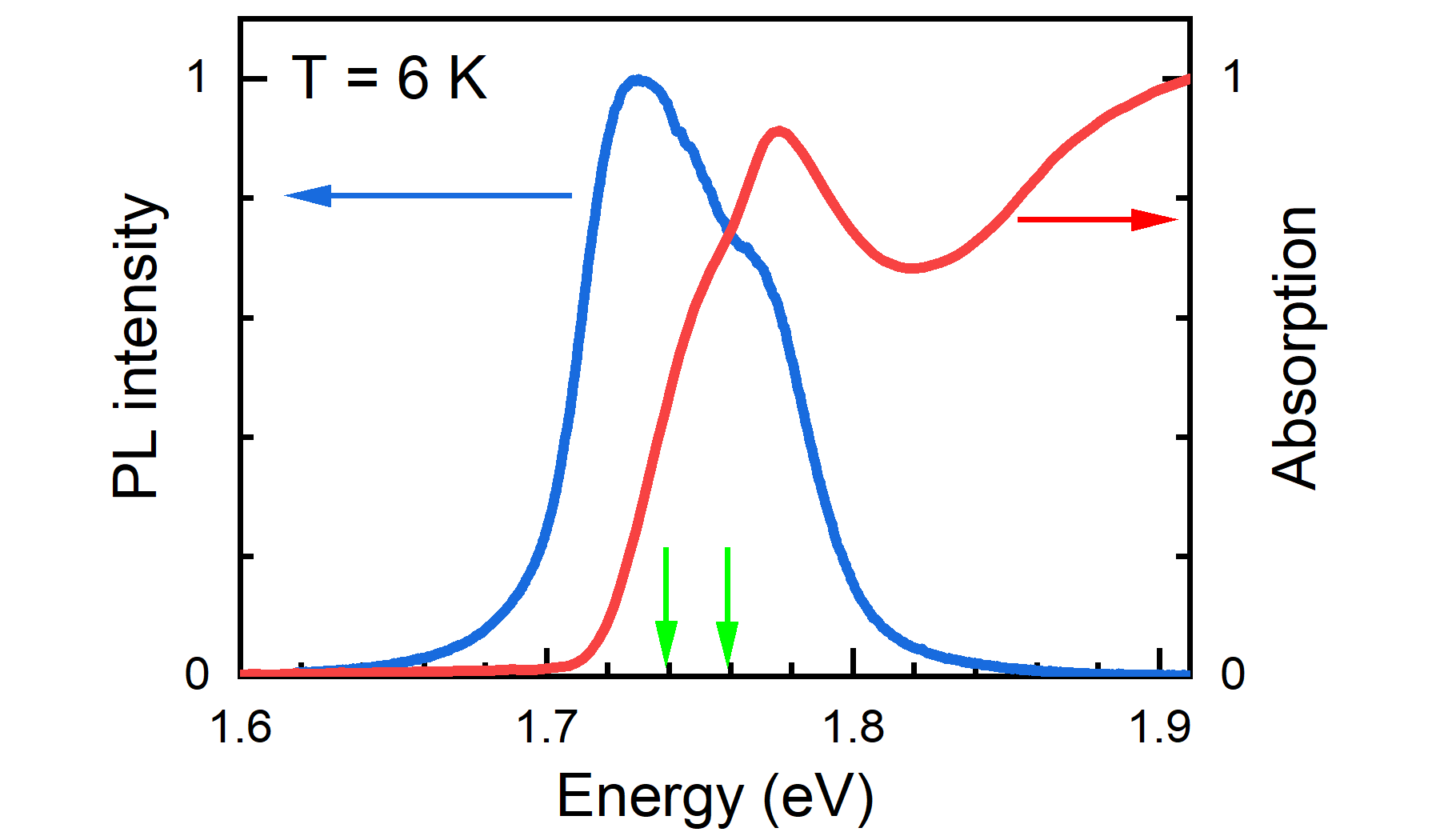}
\caption{Photoluminescence (blue line) and absorption (red line) spectra of the studied CsPbI$_3$ NCs at $T=6$~K.  Green arrows show the laser photon energies used in the pump-probe experiments. 
}
\label{fig:intro}
\end{figure}

Photoluminescence and absorption spectra of the studied CsPbI$_3$ NCs are shown in Figure~\ref{fig:intro}. The absorption spectrum shows a broad line with the maximum at $1.776$~eV and a shoulder at $1.758$~eV, which we assign to exciton absorption. The broadening is related to the rather large size dispersion of the NCs in this sample. The optical and spin properties of this sample were studied in Refs.~\cite{nestoklon2023_nl, Meliakov2024_paper1, Meliakov2025_paper2}, where it is labeled as sample~\#3. There the NC size distribution from $9$~to $12$~nm was obtained. The PL spectrum is Stokes shifted with respect to the absorption and has a comparable full width at half maximum (FWHM) of $40$~meV. It has its maximum at $1.729$~eV and a high-energy shoulder at $1.765$~eV. Such structure of the spectra evidences that the NC size distribution has two preferred values.

\subsection{Coherent spin dynamics of electrons}

In the time-resolved experiments with spectrally narrow (linewidth of about 1~meV) 1.5-ps laser pulses, a subensemble of NCs with small size distribution can be selected by resonant excitation within the PL band. We choose for this study the laser photon energy of 1.739~eV, which corresponds to the CsPbI$_3$ NCs with average size of $11$~nm. 

\begin{figure*}[hbt!]
\centering
\includegraphics[width=1.5\columnwidth]{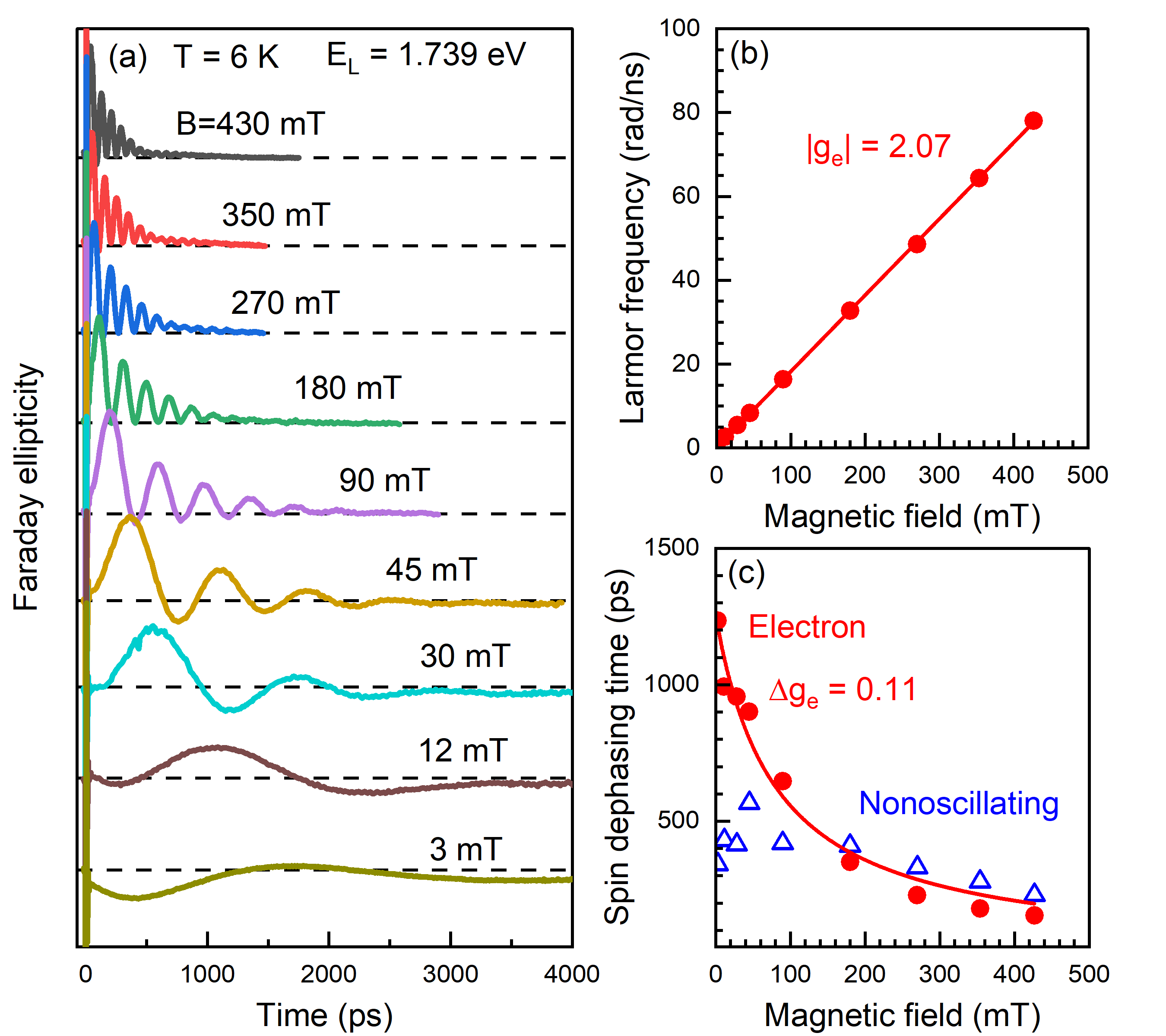}
\caption{Coherent spin dynamics of electrons in CsPbI$_3$ NCs at $T=6$~K.
(a) FE dynamics measured in Voigt magnetic fields ranging from $3$ to $430$~mT.
(b) Magnetic field dependence of the electron Larmor precession frequency. Red line is a fit with Eq.~\eqref{eq:LarmFreq} using $|g_{\rm e}|=2.07$.
(c) Magnetic field dependence of the dephasing times of the electron (red circles) and the non-oscillating (blue triangles) components. Red line is a fit with Eq.~\eqref{eq:InhDeph} using $\Delta g_{\rm e}=0.11$ and $T_\text{2,e}^*(0)=1250$~ps.
}
\label{fig:Magnetic}
\end{figure*}

Figure~\ref{fig:Magnetic}(a) shows a set of Faraday ellipticity (FE) dynamics traces, measured in different magnetic fields. One can see that in weak magnetic fields the dynamics extend over 4~ns, which is considerably longer than the exciton recombination time of  $500$~ps, as measured by time-resolved photoluminescence. This evidences that the spin signals are contributed by resident charge carriers (electrons and/or holes), whose spin lifetime is not limited by recombination. This finding is typical for lead halide perovskite NCs~\cite{Grigoryev2021,kirstein2023_SML}. The mechanism of generation of spin coherence of the resident carriers in singly-charged NCs discussed in these papers is the same as the one considered for singly-charged (In,Ga)As/GaAs quantum dots~\cite{Yakovlev_Ch6}. 

The FE dynamics in Figure~\ref{fig:Magnetic}(a) have a precessing component with frequency increasing with growing magnetic field. The oscillation amplitude is not symmetrical with respect to the zero level shown by the dashed lines, which evidences the presence of another non-oscillating component. We fit the dynamics with the following function:
\begin{equation}
A_{\rm FE}(t) \propto S_\text{0,e} \cos(\omega_\text{L}t) \exp\left(-\frac{t}{T_\text{2,e}^*}\right)
+ S_\text{no} \exp\left(-\frac{t}{\tau_{\rm no}}\right).
\label{eq:Voigt}
\end{equation}
Here, $S_\text{0,e}$ is the initial light-induced spin polarization of electrons, $T_\text{2,e}^*$ is the spin dephasing time of electrons characterizing the decay of the oscillating component, $S_\text{no}$ is the initial spin polarization corresponding to the non-oscillating component, and $\tau_{\rm no}$ is the decay time of the non-oscillating component. The
Larmor precession frequency $\omega_\text{L}$ is determined by the Land\'e $g$-factor and linearly depends on magnetic field $B$ according to:
\begin{equation}
\omega_\text{L}(B) = |g|{{\mu}_\text{B}}B/{\hbar} .
\label{eq:LarmFreq}
\end{equation}
Here, ${\mu}_\text{B}$ is the Bohr magneton and $\hbar$ is the Planck constant. Note, that Eq.~\eqref{eq:LarmFreq} does not account for possible zero-field splitting of the spin states, which are not important for relatively strong magnetic fields. We will account the zero-field splitting induced by the hyperfine interaction below in Eq.~\eqref{eq:LarmFreq1}. Fitting the experimental dynamics with  Eq.~\eqref{eq:Voigt} allows us to measure the Larmor precession frequency, the $g$-factor, the spin dephasing time $T_\text{2,e}^*$ and the decay time of the non-oscillating component.

Figure~\ref{fig:Magnetic}(b) gives the magnetic field dependence of the Larmor precession frequency. According to Eq.~\eqref{eq:LarmFreq} the slope of this linear dependence corresponds to $|g_{\rm e}|=2.07$. This value corresponds to the electron $g$-factor in CsPbI$_3$ NCs, which is expected to be positive~\cite{nestoklon2023_nl,Meliakov2025_paper2}. This allows us to assign the oscillating component to the coherent spin precession of the resident electrons.

The magnetic field dependence of the spin dephasing time of electrons ($T_\text{2,e}^*$) is shown in Figure~\ref{fig:Magnetic}(c). The $T_\text{2,e}^*$ time decreases from 1250~ps down to 160~ps with magnetic field growing from 3~mT to 430~mT. This behavior is typical for inhomogeneous spin ensembles with a finite $g$-factor spread $\Delta g_{\rm e}$. It can be described by the following expression~\cite{Yakovlev_Ch6}:
\begin{equation}
\frac{1}{T_\text{2,e}^*(B)} {\approx} \frac{1}{T_\text{2,e}^*(0)} + \frac{\Delta g_{\rm e}\mu_\text{B}B}{\hbar}.
\label{eq:InhDeph}
\end{equation}
Here, $T_\text{2,e}^*(0)$ is the spin dephasing time at zero magnetic field. Fitting the experimental data with Eq.~\eqref{eq:InhDeph} yields $\Delta g_{\rm e}=0.11$ for the electrons (the relative spread is $\Delta g_{\rm e}/g_{\rm e}=5\%$).

The blue triangles in Figure~\ref{fig:Magnetic}(c) show the magnetic field dependence of the decay time $\tau_{\rm no}$. It slowly decreases with magnetic field from about $500$~ps at $3$~mT to $230$~ps at $430$~mT. We assign the non-oscillating component to the spin dynamics of holes, which have a $g$-factor close to zero for this size of CsPbI$_3$ NCs~\cite{nestoklon2023_nl,Meliakov2025_paper2}. The other possible source of the non-oscillating component might be the decay of exciton spin polarization as result of a crystal anisotropy. This anisotropy leads to a bright exciton spin splitting. The latter is typically revealed through spin beats for circularly polarized excitation, but may lead to a monotonic decay of the spin polarization with the longitudinal exciton spin relaxation time, when the excitation contains a linear polarization component~\cite{Cai2023}. 



\subsection{Electron hyperfine interaction with nuclear spin fluctuations}

\begin{figure*}[hbt!]
\centering
\includegraphics[width=1.35\columnwidth]{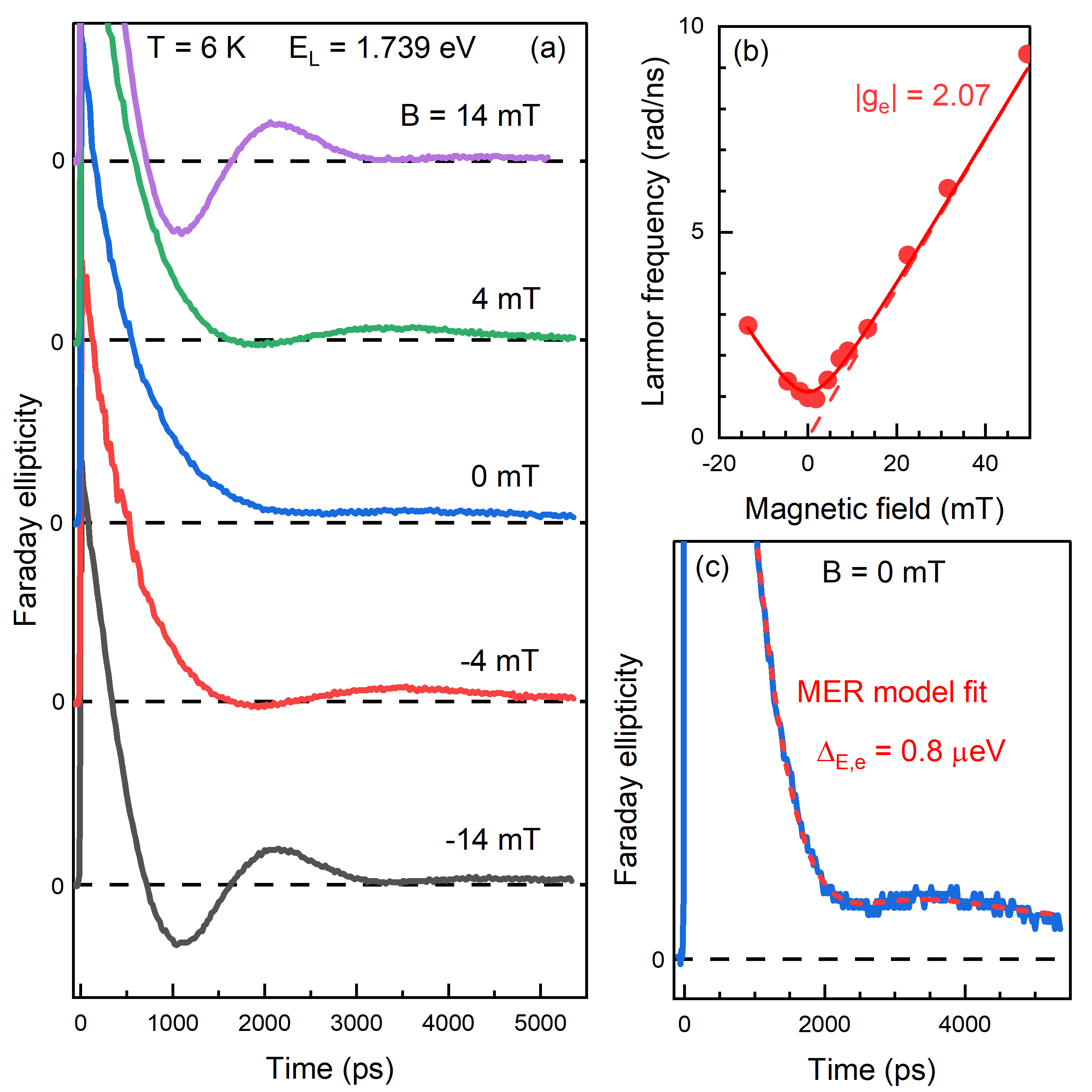}
\caption{Electron spin dynamics in  CsPbI$_3$ NCs in small Voigt magnetic fields.
(a) TRFE traces measured in different magnetic fields around zero nominal field strength from $-14$~mT to $+14$~mT. 
(b) Magnetic field dependence of the electron Larmor precession frequency. Red solid line shows fit with Eq.~\eqref{eq:LarmFreq1} with $|g_{\rm e}|=2.07$ and $\Delta_{E,{\rm e}}=0.74$~$\mu$eV. Red dashed line corresponds to Eq.~\eqref{eq:LarmFreq} with $|g_{\rm e}|=2.07$. 
(c) TRFE trace measured at zero magnetic field. Red dashed line shows fit with Eq.~\eqref{eq:MERnoField} giving $\Delta_{E,{\rm e}} = 0.8$~$\mu$eV.
}
\label{fig:Fluct}
\end{figure*}

The minimum magnetic field of 3~mT shown in Figure~\ref{fig:Magnetic}(a) is due to the remanent field of the electromagnet for the current set to zero. The magnetic field is measured with a Hall sensor. In order to study the spin dynamics in very weak magnetic fields, we tune the field with small steps in the range from $-14$~to $+14$~mT. Examples of the FE dynamics are shown in Figure~\ref{fig:Fluct}(a). We observe spin precession in all magnetic fields with the smallest Larmor precession frequency of about $1$~rad/ns at $0$~mT. 


The magnetic field dependence of the electron Larmor precession frequency evaluated from the FE dynamics is shown in Figure~\ref{fig:Fluct}(b). One can see that in weak magnetic fields it deviates from the linear dependence and does not reach zero frequency. Its minimal value of about $1$~rad/ns corresponds to an electron spin splitting of about $0.65$~$\mu$eV.
Such spin splitting can be provided by the hyperfine interaction of an electron confined in a NC with the nuclear spin fluctuations. Very recently we reported similar effect for the holes in CsPbBr$_3$ NCs~\cite{Meliakov2024_paper3}, where an energy splitting $\Delta_{E,{\rm_h}}$ of $3-5$~$\mu$eV was measured. The smaller splitting for the electrons compared to the holes is expected for lead halide perovskites, where the Bloch state specifics of the band structure in the vicinity of the band gap results in an about 5 times weaker hyperfine interaction for the electrons~\cite{kirstein2022am}.

The problem of the spin dynamics of an electron confined in a quantum dot, where it interacts with the hyperfine field of the nuclear spin fluctuations, was considered theoretically by Merkulov, Efros and Rosen (MER model) in Ref.~\onlinecite{Merkulov2002}. Their theory is developed for GaAs quantum dots, where the electron hyperfine interaction greatly exceeds the hole one. We extended MER model recently in Ref.~\onlinecite{Meliakov2024_paper3} to holes in lead halide perovskites by accounting for the difference from III-V semiconductors in crystal and band structure. Below we adopt this formalism for the electrons and use it for analysis of our experimental data. 

According to the MER model, an electron spin in a single NC precesses about the effective nuclear hyperfine magnetic field $\textbf{B}_N$ that is provided by the nuclear spin fluctuations. We refer to it here as the hyperfine nuclear field or the nuclear field. Amplitude and orientation of $\textbf{B}_N$ randomly change in time. However, the characteristic time of their changes is much longer than the electron Larmor precession period, so that one can consider the nuclear fields as ``frozen''. 

In order to model the spin dynamics in the NC ensemble, averaging over different directions and magnitudes of the hyperfine nuclear field needs to be done. The nuclear fields are distributed with the Gaussian probability density distribution with $\Delta_{B,{\rm e}}$ being the dispersion of the nuclear field distribution. If an external magnetic field $\textbf{B}$ is applied, one should consider the electron spin precession about the total field $\textbf{B}_N+\textbf{B}$, see details in Ref.~\onlinecite{Meliakov2024_paper3}. One should also take into account other mechanisms of spin dephasing and relaxation with a characteristic time $\tau_s^*$.

For zero external magnetic field the following equation for the dynamics of the projection of spin polarization onto optical axis can be obtained~\cite{Meliakov2024_paper3}:
\begin{widetext}
\begin{equation}
\left<S_z (t)\right> = \\ \frac{S_{0,\text{e}}}{3} \left\{ 1+2 \left( 1-2 \left( \frac{t}{2T_{\Delta,{\rm e}}} \right) ^2 \right) \exp \left[ - \left( \frac{t}{2T_{\Delta,{\rm e}}} \right)^2 \right] \right\}  \exp\left(- \frac{t}{\tau_s^*} \right) + S_\text{no} \exp\left(- \frac{t}{\tau_{\rm no}} \right) \,.
\label{eq:MERnoField}
\end{equation}
\end{widetext}
The decay factor $\exp\left(- t/\tau_s^*\right)$ accounts for spin dephasing mechanisms not related to the nuclear fluctuations. Also,
the second term is added to account for the non-oscillating component of the spin dynamics. $S_{0,\text{e}}$ is the initial electron spin polarization, $T_{\Delta,{\rm e}} ={T_{\rm 2,e}^*}/\sqrt{2}$ is the effective spin dephasing time of the ensemble of electron spins:
\begin{equation}
T_{\Delta,{\rm e}}= \frac{\hbar}{\Delta_{E,{\rm e}}} = \frac{\hbar}{|g_{\rm e}| \mu_B \Delta_{B,{\rm e}}}.
\label{eq:TDelta}
\end{equation}
$\Delta_{E,{\rm e}}=|g_{\rm e}| \mu_{\rm B} \Delta_{B,{\rm e}}$ is the dispersion of the electron spin splitting in the nuclear field. Fitting the dynamics at zero magnetic field (red dashed line in Figure~\ref{fig:Fluct}(c)) with Eq.~\eqref{eq:MERnoField} yields $T_{\Delta,{\rm e}} = 800$~ps and $\Delta_{E,{\rm e}} = 0.8$~$\mu$eV. 

According to the MER theory, oscillations should be observed even in zero magnetic field, which can be also described as Larmor spin precession in the hyperfine fluctuating field. In each NC, the electron spin precesses with a Larmor frequency determined by the total magnetic field $\textbf{B}_N+\textbf{B}$. Averaging over the ensemble of NCs gives the following magnetic field dependence of the Larmor frequency:
\begin{equation}
{\hbar}\omega_\text{L}(B) = \sqrt{(|g_{\rm e}|{{\mu}_\text{B}}B)^2+\Delta_{E,{\rm e}}^2}.
\label{eq:LarmFreq1}
\end{equation}
Fit of the experimental data with this equation using $|g_{\rm e}|=2.07$ gives $\Delta_{E,{\rm e}} = 0.74$~$\mu$eV (Figure~\ref{fig:Fluct}(b)), which is close to the $\Delta_{E,{\rm e}} = 0.8$~$\mu$eV obtained from the spin dynamics.

\begin{figure*}[hbt!]
\centering
\includegraphics[width=1.8\columnwidth]{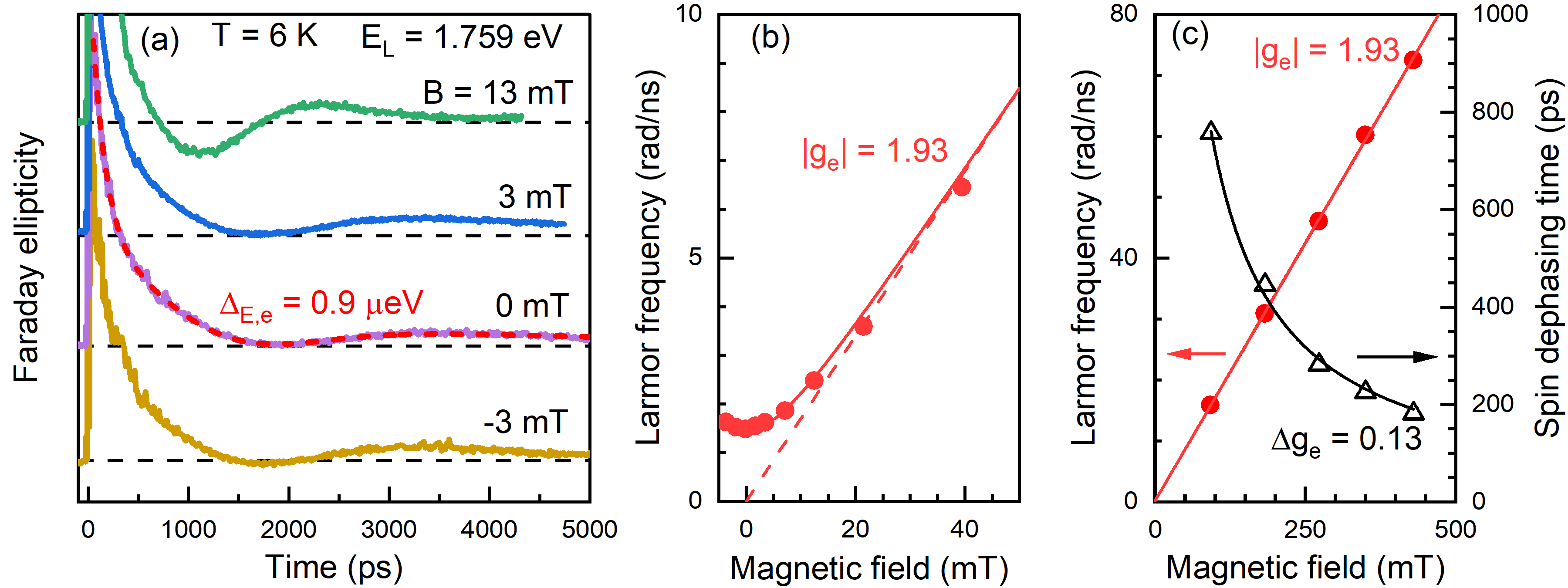}
\caption{Electron spin dynamics in CsPbI$_3$ NCs measured with femtosecond laser pulses.
(a) TRFE traces measured in different Voigt magnetic fields close to zero, varying from $-3$~mT to $+13$~mT. Red dashed line shows a fit with Eq.~\eqref{eq:MERnoField}. 
(b) Dependence of the Larmor precession frequency on magnetic field. Red solid line shows a fit with Eq.~\eqref{eq:LarmFreq1} with $|g_{\rm e}|=1.93$ and $\Delta_{E,{\rm e}}=0.95$~$\mu$eV. Red dashed line corresponds to Eq.~\eqref{eq:LarmFreq} with $|g_{\rm e}|=1.93$.
(c) Magnetic field dependences of the Larmor precession frequency (red circles) and the spin dephasing time (black triangles). Red line is a fit with Eq.~\eqref{eq:LarmFreq} using $|g_{\rm e}|=1.93$. Black line is a fit with Eq.~\eqref{eq:InhDeph} using $\Delta g_{\rm e}=0.13$.
}
\label{fig:fs}
\end{figure*}

In order to check whether the electron spin dynamics in weak magnetic fields depend on the spectral width of the laser pulse, we perform TRFE measurements with 150 fs pulses with a spectral width of about 20 meV, which is common for mode-locked laser systems used in time-resolved measurements. For this experiment we set the central laser photon energy to $E_{\rm L}=1.759$~eV. The electron spin dynamics in weak magnetic fields are shown in Figure~\ref{fig:fs}(a). The experimental features are close to what we measured with ps-pulses, see Figure~\ref{fig:Fluct}(a). At zero magnetic field the evaluated $\Delta_{E,{\rm e}} = 0.9$~$\mu$eV. We obtain a similar value of $\Delta_{E,{\rm e}} = 0.95$~$\mu$eV  from the fit of the Larmor frequency magnetic field dependence using Eq.~\eqref{eq:LarmFreq1}, as shown in Figure~\ref{fig:fs}(b). Note that these values closely match the 0.8~$\mu$eV for the ps-pulses.
The electron $g$ factor of  $|g_{\rm e}| = 1.93$  is slightly smaller at this laser photon energy, which is in line with the general trend of a decreasing $g_{\rm e}$ in smaller NCs with larger confinement energy~\cite{Meliakov2025_paper2}.

The spin dephasing times $T_\text{2,e}^*$ and their magnetic field dependences are similar for fs- and ps-pulses, compare Figures~\ref{fig:fs}(c) and~\ref{fig:Magnetic}(c). What is interesting and surprising, is that the spread of the electron $g$-factor $\Delta g_{\rm e} = 0.13$ evaluated from the magnetic field dependence of $T_\text{2,e}^*$ for fs-pulses is only slightly larger than the $\Delta g_{\rm e} = 0.11$ for ps-pulses. This evidences that $\Delta g_{\rm e}$ is determined by other mechanisms than their energy dispersion $\Delta g_{\rm e}(E)$. We suggest that variations in NC shape and orientation of their c-axis in the ensemble are responsible for that behavior.  



\section{Theory of electron-nuclear hyperfine interaction in $\rm CsPbI_3$ NCs}
\subsection{Electron spin splitting in fluctuating nuclear fields}

In this section we derive the equation connecting the experimentally measured hyperfine energy dispersion $\Delta_{E,{\rm e}}$ with the size of the NCs, the hyperfine interaction constants, and the parameters of the band structure.

The Bloch amplitudes of the electrons and the holes at the bottom of conduction band and the top of valence band in lead halide perovskites are strongly contributed by linear combinations of $s$- and $p$-type atomic orbitals of the lead and the halogens, which determine the hyperfine interaction with corresponding nuclear spins. There is also considerable admixture of other states, whose coupling with nuclear spins, however, is much less effective. 
Taking into account the spin-orbit interaction, in line with the Supplementary Material of Ref.~\cite{kirstein2022am}, one arrives at the following expressions for the $p$-type Bloch amplitudes at the bottom of the conduction band:
\begin{equation}
\begin{aligned}
u_{cbb}^{+1/2} &= -\frac{ \ket{Z_{cb}}\uparrow + \left( \ket{X_{cb}} + i \ket{Y_{cb}} \right)\downarrow }{\sqrt{3}}\,, \\
u_{cbb}^{-1/2} &= \frac{\ket{Z_{cb}}\downarrow - \left( \ket{X_{cb}} - i \ket{Y_{cb}} \right)\uparrow}{\sqrt{3}}\,.
\end{aligned}
\label{eq:KK1}
\end{equation}
Here the indices $\pm 1/2$ denote the $Z$-projections of the effective spin of the conduction band quasiparticles (electrons), $\uparrow$ and $\downarrow$ are the spinor components of the actual electron with the orbital functions
\begin{equation}
\begin{aligned}
\left| X_{cb} \right\rangle &= C_{Pb}^{p} \left| X_{Pb} \right\rangle + C_{hal}^{s} \left| S_{halX} \right\rangle + C_{res} \left| X_{res} \right\rangle \\
\left| Y_{cb} \right\rangle &= C_{Pb}^{p} \left| Y_{Pb} \right\rangle + C_{hal}^{s} \left| S_{halY} \right\rangle + C_{res} \left| Y_{res} \right\rangle \\ 
\left| Z_{cb} \right\rangle &= C_{Pb}^{p} \left| Z_{Pb} \right\rangle + C_{hal}^{s} \left| S_{halZ} \right\rangle + C_{res} \left| Z_{res} \right\rangle,
\end{aligned}
\label{eq:KK2}
\end{equation}
where $\left| X_{Pb} \right\rangle$, $\left| Y_{Pb} \right\rangle$, and $\left| Z_{Pb} \right\rangle$ are the $p$-type atomic orbitals of lead, while $\left| S_{halX} \right\rangle$, $\left| S_{halY} \right\rangle$, and $\left| S_{halZ} \right\rangle$ are the atomic $s$-orbitals of the halogen atoms located correspondingly along the $X$, $Y$ and $Z$ cubic axes from the lead atom. $\left| X_{res} \right\rangle$, $\left| Y_{res} \right\rangle$, and $\left| Z_{res} \right\rangle$ are the residual parts of the $p$-type Bloch amplitudes, which do not contribute considerably to the hyperfine interaction. The coefficients $C$ are defined such that they provide the normalization of the Bloch amplitude to unity: $\left| C^{p}_{Pb} \right|^{2} + \left| C^{s}_{hal} \right|^{2} + \left| C_{res} \right|^{2} = 1$.

In CsPbI$_3$ NCs the hyperfine interaction of the electron at the bottom of the conduction band is contributed by the lead and iodine isotopes with nonzero spins. Note that the contribution of the cesium orbitals to the electron states at this energy is negligibly small. 


By calculating the matrix elements of the Hamiltonian of the hyperfine interaction \cite{Abragam1961}
\begin{equation}
\hat{H}_{hf} = 2 \mu_{\rm B} \hbar \gamma_N \mathbf{I} 
\left[ 
\frac{\mathbf{l}}{r^3} - \frac{\mathbf{s}}{r^3} + 3 \frac{\mathbf{r} (\mathbf{s} \cdot \mathbf{r})}{r^5} + \frac{8}{3} \pi \mathbf{s} \delta(\mathbf{r})
\right]
\label{eq:KK3}
\end{equation}
on these basis states (in Eq.~\eqref{eq:KK3} $\mathbf{l}$, $\mathbf{s}$, $\mathbf{r}$ are orbital momentum, spin and position of the electron, respectively),  one arrives to the following spin Hamiltonian for the hyperfine interaction in the conduction band
\begin{multline}
\hat{H}^{c}_{eN} = \sum_i v_0 \left| \Psi^2_e (\mathbf{R}_i) \right|^2 ( A^{cb}_{Pb} \mathbf{I}_{Pb,i} + \mathbf{I}_{halX,i} \cdot \hat{A}^{cb}_{halX} \\
+ \mathbf{I}_{halY,i} \cdot \hat{A}^{cb}_{halY}
+ \mathbf{I}_{halZ,i} \cdot \hat{A}^{cb}_{halZ} ) 
\mathbf{S}_e.
\label{eq:KK4}
\end{multline}
Here $v_0$ is the unit cell volume, $\mathbf{S}_e$ is a total angular momentum of the electron in conduction band, $\left| \Psi^{2}_{e} (\mathbf{R}_{i}) \right|^{2}$ is the probability density of the envelope function of the conduction-band electron, $i$ enumerates the elementary cells, $\mathbf{I}_{Pb,i}$, $\mathbf{I}_{halX,i}$, $\mathbf{I}_{halY,i}$, and $\mathbf{I}_{halZ,i}$ are the nuclear spin operators of lead and the halogens situated along the $X$, $Y$, $Z$ axis with respect to the lead nucleus. The hyperfine interaction of the electron with the lead nuclear spin is isotropic and is characterized by the scalar constant:
\begin{equation}
A^{cb}_{Pb} = A^{p}_{Pb} \left| C^{p}_{Pb} \right|^2,
\label{eq:KK5}
\end{equation}
where $A_{Pb}^p$ is the atomic hyperfine constant for the outer-shell $p$-electron of lead. The hyperfine interaction of the electron with the halogens is anisotropic. It depends on the position of the halogen nucleus, and is characterized by the tensors
\begin{subequations}\label{eq:KK6}
\begin{align}
 \hat{A}^{cb}_{halX} & = \frac{\left| C^{s}_{hal} \right|^2}{3} A^{s}_{hal} 
    \begin{pmatrix}
        1 & 0 & 0 \\
        0 & -1 & 0 \\
        0 & 0 & -1
    \end{pmatrix} \,,
\\
\hat{A}^{cb}_{halY} &= \frac{\left| C^{s}_{hal} \right|^2}{3} A^{s}_{hal} 
    \begin{pmatrix}
        -1 & 0 & 0 \\
        0 & 1 & 0 \\
        0 & 0 & -1
    \end{pmatrix} \,,
\\
\hat{A}^{cb}_{halZ} &= \frac{\left| C^{s}_{hal} \right|^2}{3} A^{s}_{hal} 
    \begin{pmatrix}
        -1 & 0 & 0 \\
        0 & -1 & 0 \\
        0 & 0 & 1
    \end{pmatrix},
\end{align}
\end{subequations}
where $A_{hal}^s$ is the atomic hyperfine constant for the outer-shell $s$-electron of the halogen.

Now, for the mean squared spin splitting of a localized conduction-band electron in the random hyperfine field of unpolarized nuclear spins (equal to the squared hyperfine energy dispersion) we get
\begin{widetext}
\begin{multline}
{\Delta_{E,{\rm e}}}^2 = \left\langle \sum_i \left[ v_0 \left| \Psi^2_e (\mathbf{R}_i) \right|^2 \left( A^{cb}_{Pb} \mathbf{I}_{Pb,i} + \mathbf{I}_{halX,i}\hat{A}^{cb}_{halX} + \mathbf{I}_{halY,i} \hat{A}^{cb}_{halY} +  \mathbf{I}_{halZ,i} \hat{A}^{cb}_{halZ} \right) \right]^{2}  \right\rangle = \\
 = v_0^2  \sum_i  \left|  \Psi^2_e (\mathbf{R}_i)   \right|^{4} \left[ I_{Pb}(I_{Pb}+1)\alpha_{Pb}N_{Pb}\left( A^{p}_{Pb} | C^{p}_{Pb}|^{2}\right)^{2} + I_{hal} (I_{hal}+1)\alpha_{hal}N_{hal}\left( A^{s}_{hal}\frac{\left| C^{s}_{hal}\right|^2}{3}\right)^{2}\right]  \, ,
\label{eq:KK7}
\end{multline}
\end{widetext}
where $\alpha_{Pb}$ and $\alpha_{hal}$ are the abundances of the magnetic isotopes of lead and halogen, $N_{Pb} = 1$ and $N_{hal} = 3$ are the numbers of the corresponding nuclei in the unit cell.

If the number of cells in the nanocrystal is large, the sum in Eq.~\eqref{eq:KK7} can be replaced with an integral:
\begin{equation}
v_0^2 \sum_i \left| \Psi_e (\mathbf{R}_i) \right|^{4} \approx v_0 \int \left| \Psi^2_e ({\bf R}) \right|^{4} {\rm d}^3 {\bf R} \, .
\label{eq:KK8}
\end{equation}

For a spherical CsPbI$_3$ nanocrystal with iodine ($I_{hal} = 5/2$, $\alpha_{hal} = 1$) as halogen, we obtain
\begin{widetext}
\begin{equation}
 \Delta_{E,{\rm e}}^2 = \frac{a_0^3}{0.28 d^3} \left[ \frac{3}{4} \alpha_{Pb} (A^{p}_{Pb})^2 |C^{p}_{Pb}|^4 + \frac{35}{12} (A^{s}_{I})^2 |C^{s}_{I}|^4 \right] \, ,
\label{eq:Hyperfine}
\end{equation}
\end{widetext}
where $d$ is the nanocrystal diameter and $a_0=0.624$~nm is the CsPbI$_3$ lattice constant. $A^{s}_{I}$ is the hyperfine constant of the iodine halogen $A^{s}_{hal}$. $C^{s}_{I}$ is the contribution of the iodine halogen $s$-orbitals to the bottom of conduction band.

According to the available literature, $A^p_{Pb}$ ranges from $18.2 \, \mu$eV (Ref.~\onlinecite{Morton1978}, Hartree-Fock calculations) to $50.4 \, \mu$eV (Ref.~\onlinecite{Hewes1973}, Knight field measurement in PbTe). For iodine, $A^s_I \approx 264\pm10 \, \mu$eV was determined from infrared spectra of atomic iodine \cite{Luc-Koenig1975}. A theoretical value, obtained in the same work, is almost two times smaller amounting to $135.5 \, \mu$eV, while in the tables by Morton and Preston \cite{Morton1978} a larger calculated value of $171.3 \, \mu$eV is given.

\subsection{DFT calculations of Pb and I orbital contributions to the electron hyperfine interaction}


\begin{figure*}
  \includegraphics[width=0.8\linewidth]{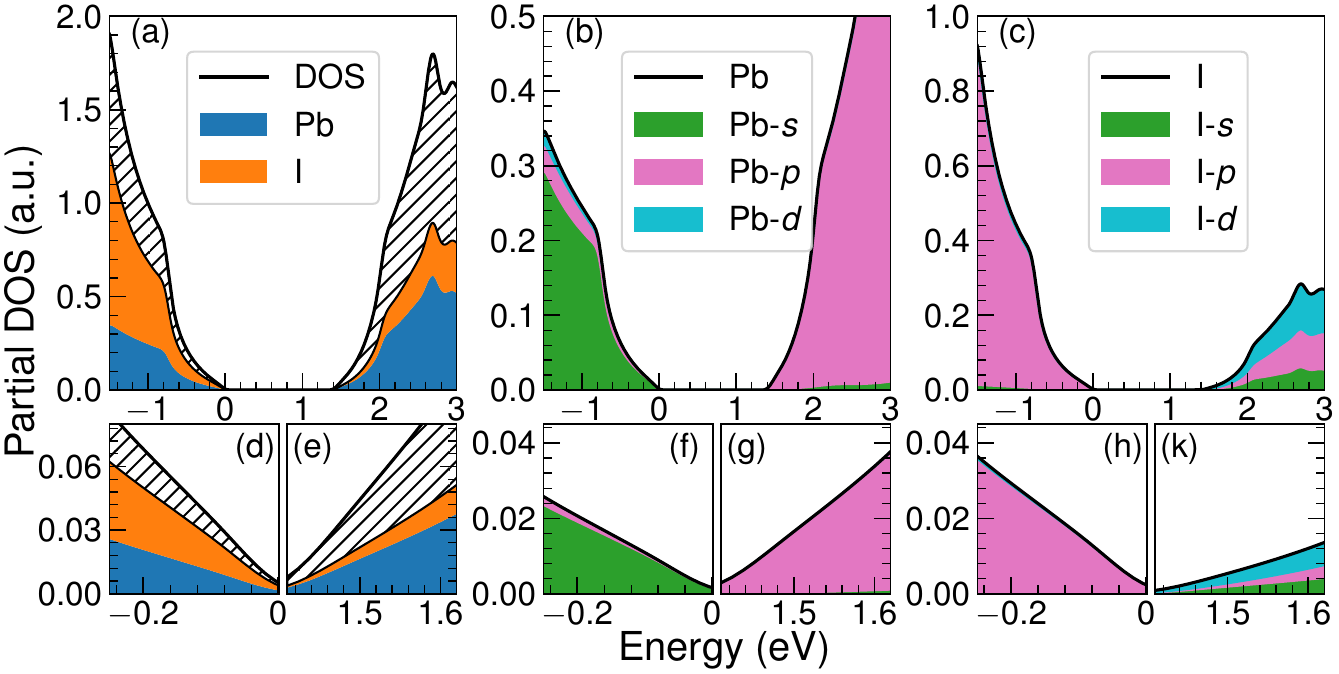}
 \caption{
PDOS for cubic CsPbI\textsubscript{3} bulk crystals calculated in DFT. (a) Total DOS (solid line) and PDOS on the Pb and I atoms (filled areas). The difference between the sum of PDOS on atoms and the total DOS gives the PDOS at distances more than $R_{MT}$ from the atomic sites  (dashed area). 
 (b,c) PDOS on the Pb and I atoms resolved according to their angular momentum: black line shows the total DOS on atoms (shown by filled areas in panel (a)) and filled areas give the contribution from $s$- (angular momentum 0), $p$- (angular momentum 1), and $d$- orbitals (angular momentum 2) to the PDOS on atoms. (f-k) Zooms of panels (b,c) for the states in vicinity of the band gap. 
}\label{fig:pDOS}
\end{figure*}

As we can see from Eq.~\eqref{eq:Hyperfine}, the hyperfine energy dispersion depends on the contributions of Pb- and I-orbitals to the bottom of the conduction band states. We calculate these contributions using density functional theory (DFT).
The DFT calculations are done with the package WIEN2k \cite{Blaha2020}. Since this is a full-potential code, which does not use pseudopotentials, it gives reasonable values for the wave functions near the nuclei. The calculations are made for a CsPbI\textsubscript{3} bulk crystal in the cubic phase with the lattice constant of $a_0=0.624$~nm. In the calculations, the ``automatic'' precision setting \verb|-prec 3n| with fine $k$ mesh $24\times 24\times 24$ is used. The accurate analysis of the convergence shows that the convergence of mBJ is coarse compared to the standard GGA exchange-correlation functionals \cite{Nestoklon2021}. The energy is converged up to $10^{-5}$ a.u. With these parameters we obtain the band gap value of $E_g=1.408$~eV. All calculations are done with accounting for the spin-orbit interaction.

In WIEN2k, the Kohn-Sham equations are solved in the basis, which is obtained as combination of Slater-like numerical orbitals within the spheres with radius $R_{MT}$ centered at the atomic cites and as plane waves between these spheres. This allows one to analyze the atom- and angular-momentum-resolved density of states (DOS).

The partial density of states (PDOS) near the band gap is shown in Figure~\ref{fig:pDOS}. Here, zero energy corresponds to the top of the valence band. In these calculations $R_{MT}=2.45$ for the Cs atoms, $R_{MT}=2.55$ for the Pb atoms, and $R_{MT}=2.5$ for the I atoms are used. These values are chosen somewhat larger than the standard settings to minimize the ratio of the PDOS between the spheres around the atoms (``interatomic'' DOS), even though the estimation of {$|C_{Pb/I}^{\zeta}|$ ($\zeta=s,p$)}  should not depend on the choice of $R_{MT}$ to be considered as physically relevant. It is instructive to discuss first the ratio of the DOS on atoms and next on different orbitals of the atoms. In Figure~\ref{fig:pDOS} the atom-resolved DOS, the orbital-resolved PDOS on Pb and the orbital-resolved DOS on I are shown. One can see that the valence band is dominated by $p$-orbitals of I and $s$-orbitals of Pb, while the conduction band is mostly built from $p$-orbitals of Pb with a small admixture of $s$-orbitals of I. The calculated contribution of the Cs DOS near the bottom of the conduction band is negligible.

\begin{figure}
  \includegraphics[width=1\linewidth]{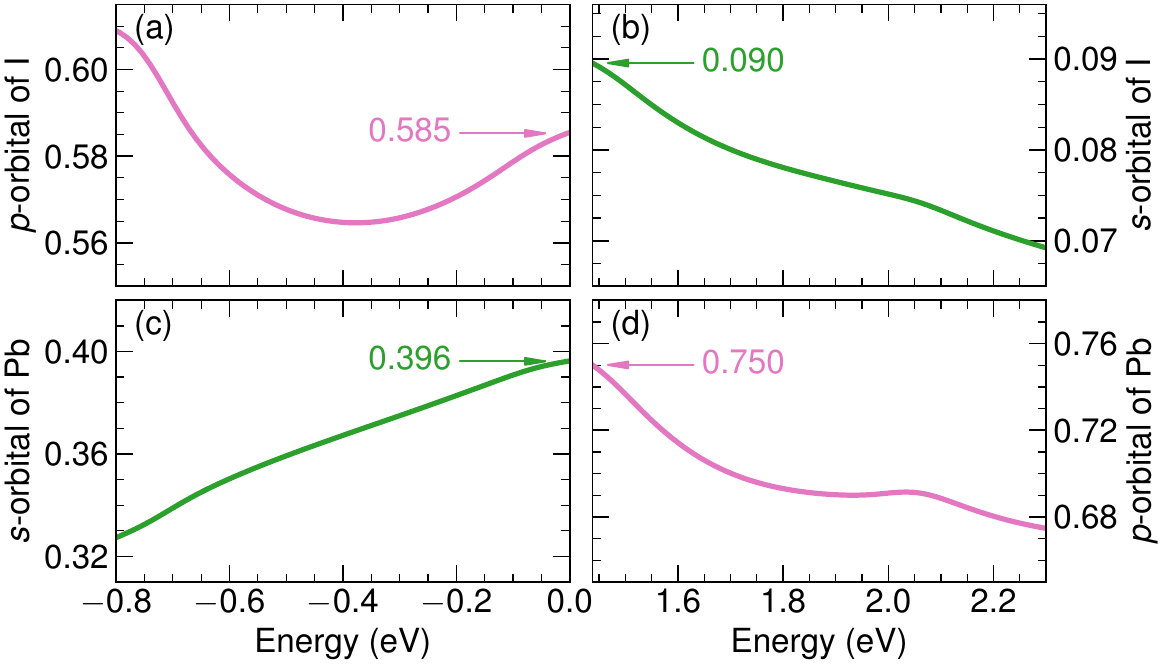}
 \caption{
Ratios of angular momentum-resolved partial DOS of iodide and lead orbitals normalized to the total PDOS on atoms in a cubic CsPbI\textsubscript{3} bulk crystal calculated in DFT. Ratio of the (a) $p$-orbitals of I and (c) $s$-orbitals of Pb in the valence band. Ratio of the (b) $s$-orbitals of I and (d) $p$-orbitals of Pb in the conduction band. 
}\label{fig:pDOS_orbitals}
\end{figure}

For deeper insight into the compositions of the states in vicinity of the band gap, in Figure~\ref{fig:pDOS_orbitals} we show the ratios of the DOS of interest in the valence and conduction bands defined as 
\begin{equation}
|C_{j}^{\zeta}|^2 = \frac{{\rm PDOS}(j, \zeta)}{\sum_{\xi} {\rm PDOS}(Pb, \xi) + {\rm PDOS}(I, \xi)}\,,
\end{equation}
where ${\rm PDOS}(j, \zeta)$ is the partial DOS at $j$-th atom with angular momentum $\zeta$.
As can be seen,  the conduction band bottom is composed mostly of lead $p$-orbitals which contribute with 75\% and iodine $s$-orbitals which contribute with 9\%. For completeness, we note that the top of the valence band is composed from 58.5\% $p$-orbitals of I and $39.6$\% of Pb $s$-orbitals. Note that these numbers are stable with respect to the value of $R_{MT}$. For $R_{MT}=2.35$ (on all atoms) these values are 72.7\%/10.9\% ($p$-Pb/$s$-I in the conduction band) and 58\%/40.7\% ($p$-I/$s$-Pb in the valence band). Use of the standard PBE potential \cite{Perdew1996} gives 62.6\%/6.7\% in the conduction band and 63.8\%/29.8\% in the valence band, while severely underestimating the band gap ($E_g^{\rm PBE}=0.316$~eV).

\section{Discussion}

\begin{figure*}[hbt!]
\includegraphics[width=1.5\columnwidth]{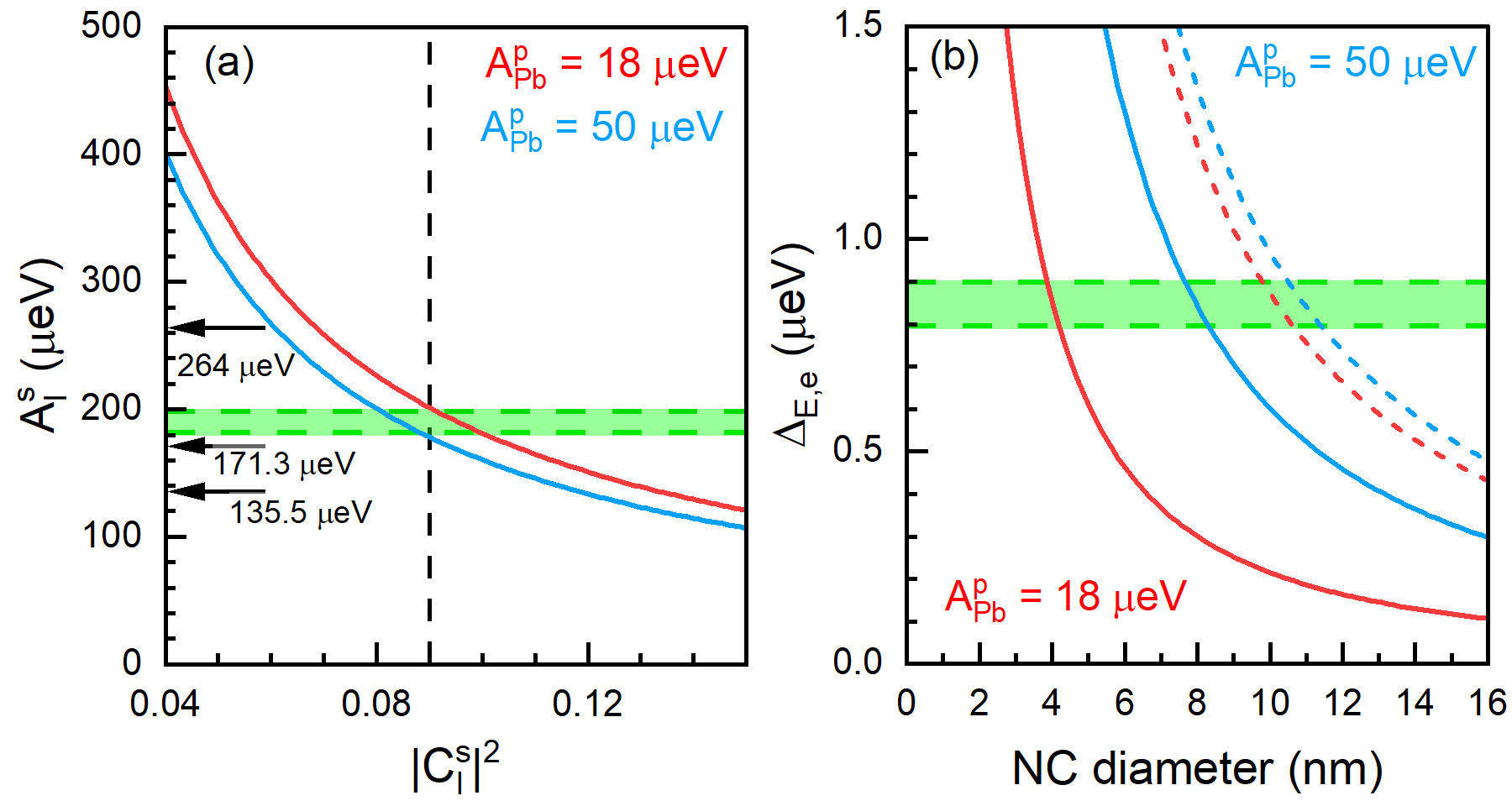}
\caption{
Evaluation of the parameters responsible for the spin dynamics of electrons interacting with the lead and iodine spins in CsPbI$_3$ NCs. Red lines correspond to $A_{Pb}^p = 18$~$\mu$eV, blue lines correspond to $A_{Pb}^p = 50$~$\mu$eV. 
(a) Parameters $A_{I}^s$ (vertical axis) and $|C_{I}^{s}|^2$ (horizontal axis), which satisfy Eq.~\eqref{eq:Hyperfine} for the experimental value of $\Delta_{E,{\rm e}} = 0.8$~$\mu$eV and $|C_{Pb}^{p}|^2=0.75$ from DFT calculations. Vertical dashed line indicates the value of $|C_{I}^{s}|^2=0.09$ from DFT calculations, green area shows the corresponding range of the hyperfine constant $A_{I}^s$. Vertical arrows give the values of $A_{I}^s$ of 135.5~$\mu$eV, 171.3~$\mu$eV, and 264~$\mu$eV from Refs. \cite{Morton1978, Luc-Koenig1975}. 
(b) NC size dependence of the cumulative contributions of the lead and iodine isotopes to $\Delta_{E,{\rm e}}$, calculated with Eq.~\eqref{eq:Hyperfine} for $|C_{Pb}^{p}|^2=0.75$ and $|C_{I}^{s}|^2=0$ (solid lines) or $|C_{I}^{s}|^2=0.09$ (dashed lines). $A_{I}^s$  is taken to be  190~$\mu$eV from the green area in panel (a). Green dashed line shows the range of experimental values of the dispersion $\Delta_{E,{\rm e}} = 0.8-0.9$~$\mu$eV measured in this work.
}
\label{fig:theory}
\end{figure*}
 
The presented experimental and theoretical results allow us to suggest an approach for evaluation of the parameters of the hyperfine interaction in the conduction band of perovskite semiconductors. We use Eq.~\eqref{eq:Hyperfine} to model the contributions of the lead and iodine isotopes to the electron hyperfine dispersion $\Delta_{E,{\rm e}}$ in the studied NCs with a medium size of 11~nm. Accounting only for the lead contribution even for $|C^{p}_{Pb}|^2 = 1$ gives $\Delta_{E,{\rm e}}=0.2-0.5$~$\mu$eV, which is insufficient to describe the experimental data. Thus, we should also account for the electron hyperfine interaction with the iodine nuclei. 

Figure~\ref{fig:theory}(a) shows the values of the hyperfine interaction constant $A_{I}^s$ and the iodine contribution $|C^{s}_{I}|^2$ which satisfy Eq.~\eqref{eq:Hyperfine} for the experimentally measured dispersion of $\Delta_{E,{\rm e}} = 0.8$~$\mu$eV. The red and blue curves correspond to $A_{Pb}^p =18$~$\mu$eV and $50$~$\mu$eV, respectively. $|C^{p}_{Pb}|^2 = 0.75$ is taken from the DFT calculations. The blue and red lines in Figure~\ref{fig:theory}(a) are close to each other, which means that the hyperfine interaction with the iodine nuclei is dominant. By taking $|C^{s}_{I}|^2=0.09$ from the DFT calculations, we determine $A_{I}^s \approx 190$~$\mu$eV, which falls into the range of previously reported values \cite{Luc-Koenig1975, Morton1978} (shown with the arrows in Figure~\ref{fig:theory}(a)). Our value of $A_{I}^s$ is about 30\% smaller than the experimental constant obtained by atomic spectroscopy in Ref.~\onlinecite{Luc-Koenig1975}, which may result from different normalization conditions for the outer electron shells in a crystal, as compared to a free atom.  
If, reversely, $A_{I}^s$ is set equal to the spectroscopic value of 264 $\mu$eV, $|C^{s}_{I}|^2$ appears to be in the range from 0.06 to 0.07, which is also in reasonable agreement with the DFT prediction.

Using $A_{I}^s \approx 190$~$\mu$eV obtained from Figure~\ref{fig:theory}(a), as well as $|C^{p}_{Pb}|^2 = 0.75$ and $|C^{s}_{I}|^2=0.09$ from DFT calculations, we evaluate how the dispersion $\Delta_{E,{\rm e}}$ depends on the NC size. The solid lines in Figure~\ref{fig:theory}(b) show only the lead contribution to $\Delta_{E,{\rm e}}$, while the dashed lines show the sum of the lead and iodine contributions. One can see that in case of the electrons in CsPbI$_3$ NCs, the hyperfine interaction with the iodine nuclei dominates over interaction with the lead nuclei even for small $|C^{s}_{I}|^2$.

\section{Conclusions}

To summarize, we investigate the coherent spin dynamics of electrons in CsPbI$_3$ perovskite nanocrystals using the time-resolved Faraday ellipticity. In high magnetic fields the electron spin dephasing is dominated by the spread of $g$-factors, which is independent of the spectral width of the pump laser pulses. In magnetic fields $B < 10$~mT, we observe electron spin precession and dephasing in the hyperfine field of the nuclear spin fluctuations. We measure the dispersion of the hyperfine interaction energy fluctuations of $\Delta_{E,{\rm e}} = 0.8-0.9$~$\mu$eV. Our model analysis shows that the electron hyperfine interaction is dominated by the interaction with the iodine nuclei. Comparison of the model analysis with the results of DFT calculations allows us to estimate the iodine hyperfine interaction constant of $A_{I}^s \approx 190$~$\mu$eV.


Our experimental and theoretical results highlight the role of the halogen nuclear spins for the electron spin dynamics in lead halide perovskite NCs. There is an important difference between the Pb$^{207}$ isotope with spin $1/2$ and the halogen isotopes with spins larger than $1/2$ (in particular, iodine with spin $5/2$). Namely, quadrupole interactions are present in the system of iodine spins. This opens a way to effects known for conventional III-V and II-VI semiconductors, but have not been investigated so far for the lead halide perovskites. One of the bright examples here is the recent discovery of long-term periodical oscillations in the nuclear spin system of (In,Ga)As that can be described as the behavior of a continuous time crystal~\cite{Greilich2025}.

\section*{Acknowledgments}
We acknowledge fruitful discussions with D. S. Smirnov and M. M. Glazov. Research performed at the P. N. Lebedev Physical Institute was financially supported by the Ministry of Science and Higher Education of the Russian Federation, Contract No. 075-15-2021-598.   E.V.K., K.V.K. and M.S.K. acknowledge support by the Saint Petersburg State University (Grant No. 125022803069-4). 
M.O.N acknowledges the financial support by the Deutsche Forschungsgemeinschaft (project AK40/13-1, no. 506623857) and the computing time provided on the Linux HPC cluster at TU Dortmund (LiDO3), partially funded in the course of the Large-Scale Equipment Initiative by the DFG as project 271512359.

\setcounter{equation}{0}
\setcounter{figure}{0}
\setcounter{table}{0}
\setcounter{section}{0}
\setcounter{subsection}{0}
\renewcommand{\theequation}{A\arabic{equation}}
\renewcommand{\thefigure}{A\arabic{figure}}
\renewcommand{\thetable}{A\arabic{table}}
\renewcommand{\thesubsection}{A\arabic{subsection}}







%

\textbf{AUTHOR INFORMATION}

{\bf Corresponding Authors} \\
Sergey~R.~Meliakov,  Email: melyakovs@lebedev.ru   \\
Dmitri R. Yakovlev,  Email: dmitri.yakovlev@tu-dortmund.de\\

\textbf{ORCID}\\
Sergey~R.~Meliakov         0000-0003-3277-9357 \\  
Evgeny~A.~Zhukov:          0000-0003-0695-0093 \\  
Vasilii~V.~Belykh:         0000-0002-0032-748X \\
Kirill V. Kavokin:         0000-0002-0047-5706 \\
Mikhail O. Nestoklon:      0000-0002-0454-342X \\
Elena V. Kolobkova:        0000-0002-0134-8434 \\  
Maria S. Kuznetsova:       0000-0003-3836-1250 \\  
Manfred Bayer:             0000-0002-0893-5949 \\
Dmitri R. Yakovlev:        0000-0001-7349-2745 \\  


\clearpage

\setcounter{equation}{0}
\setcounter{figure}{0}
\setcounter{table}{0}
\setcounter{section}{0}
\setcounter{subsection}{0}
\setcounter{page}{1}
\renewcommand{\theequation}{S\arabic{equation}}
\renewcommand{\thefigure}{S\arabic{figure}}
\renewcommand{\thepage}{S\arabic{page}}
\renewcommand{\thetable}{S\arabic{table}}
\renewcommand{\thesubsection}{S\arabic{subsection}}

\end{document}